\documentstyle[psfig,epsfig]{mn}

\newcommand{\be}{\begin{equation}}
\newcommand{\ee}{\end{equation}}
\newcommand{\ba}{\begin{eqnarray}}
\newcommand{\ea}{\end{eqnarray}}

\newcommand{\C}{\mbox{\boldmath $C$}}

\newcommand{\mub}{\mbox{\boldmath $\mu$}}
\newcommand{\Pb}{\mbox{\boldmath $P$}}

\newcommand{\n}{\mbox{\boldmath $n$}}

\newcommand{\Pp}{\mbox{\boldmath $P$}}

\newcommand{\rgl}{\rangle}
\newcommand{\lgl}{\langle}

\newcommand{\Tr}{{\rm Tr}\,}
\newcommand{\half}{\frac{1}{2}}

\newcommand{\calL}{\mbox{${\cal L}$}}

\newcommand{\mnras}{MNRAS}
\newcommand{\prd}{Phys. Rev. D}
\newcommand{\apjl}{Astro. Phys. Journal Letters}
\newcommand{\apj}{Astro. Phys. Journal}
\newcommand{\aap}{AAP}

\def\gs{\mathrel{\raise1.16pt\hbox{$>$}\kern-7.0pt %
\lower3.06pt\hbox{{$\scriptstyle \sim$}}}}         %
\def\ls{\mathrel{\raise1.16pt\hbox{$<$}\kern-7.0pt %
\lower3.06pt\hbox{{$\scriptstyle \sim$}}}}         %

\title[Mitigation of NonLinear Uncertainity]{On Mitigation of the Uncertainty in Nonlinear Matter 
Clustering for Cosmic Shear Tomography}

\author[T. D. Kitching \& A. N. Taylor]
       {T. D. Kitching\thanks{tdk@roe.ac.uk}
\& A. N. Taylor\\
SUPA, Institute for Astronomy, University of
Edinburgh, Royal Observatory, Blackford Hill, Edinburgh, EH9 3HJ,
U.K.}

\date{}

\pagerange{\pageref{firstpage}--\pageref{lastpage}}

\pubyear{2010}

\begin{document}

\maketitle

\label{firstpage}

\begin{abstract}
We present a new method that deals with the uncertainty in matter-clustering 
in cosmic shear power spectrum analysis that arises mainly due to 
poorly understood nonlinear baryonic processes on small-scales. 
We show that the majority of information about dark energy physics 
contained in the shear power comes from these small-scales; 
removing these nonlinear scales from a cosmic shear
analysis results  
in a 50\% cut in the accuracy of measurements of  dark energy
parameters, marginalizing over all other parameters. In this paper we
propose a method to recover the information on small-scales by
allowing  
cosmic shear surveys to measure the nonlinear matter power spectrum
themselves and marginalize over all possible power spectra using path
integrals. Information is still recoverable in these nonlinear regimes
from the geometric part of weak lensing. In this self-calibration
regime we find we recover 90\% of the information on dark
energy. 
Including an informative prior, we find the nonlinear matter power 
spectrum needs to be accurately known to 1\%  
down to $k=50 h$Mpc$^{-1}$ to recover 99\% of the dark energy information. 
This presents a significant theoretical challenge to understand baryonic 
effects on the scale of galaxy haloes. 
However self-calibration from weak lensing may also 
provide observational input to help constrain baryon physics.
\end{abstract}


\section{Introduction}

Cosmic shear has been identified as being a particularly sensitive tool
in understanding dark energy (Albrecht et al., 2006; Peacock et al., 2006), 
dark matter (Massey, Kitching, Richards, 2010), neutrino
physics (Hannestaad, 2010) and potential depatures from general relativity. 
There are a number of on-going (CFHTLenS, Pan-STARRS) and planned
experiments (KIDS, DES, LSST, Euclid) whose primary scientific goals
are to use cosmic shear to constrain cosmological parameters. 

However, as we show in this article, the majority of dark energy
information comes from small scales, that are expected to be
influenced by poorly understood non-linear effects such as baryonic
processes (see White, 2004; Zhan \& Knox, 2004, Mead et al., 2010; Rudd
et al., 2008; Guillet et al., 2010 for example); which may be very
difficult to model to sufficient accuracy. Indeed we may  not even
know the sign of contribution from 
baryonic effects. If dissipation is the main effect,
we can imagine baryonic collapse will lead to an enhancement of matter
clustering, or Baryonic Compression. But nonlinear feedback from star
or AGN production could also blow out baryons, reducing the overall
mass and allowing the dark matter to disperse. The formation of stars
or AGN in simulations is usually governed by sub-grid semi-analytic prescriptions,
which makes the understanding of such processes uncertain. As a step forward it would be
very useful to even quantify our uncertainty on the physics involved
on different scales.  

In addition to the uncertain physical effects there 
is the practical consideration of the accuracy of fitting formulae.  
The matter power spectrum as a function of redshift 
can be computed using linear perturbation theory of the underlying
initial density field, but by comparison
with N-body simulations it is apparent that at scales where the
matter overdensity $\delta$ becomes greater than $0.1$ 
the linear theory predictions
cannot be used as non-linear effects in the growth of structure 
become dominant. 
The most widely used corrections are Smith et al. (2003) {\tt
  halofit}, Peacock \& Dodds (1996) and the {\tt Coyote} 
formula Heitmann, et al, (2010);
{\tt halofit} and {\tt Coyote} are accurate to approximately $5-10\%$. 
In addition to this uncertainty the formula are proposed by only sampling 
a small discrete number of points in parameter space, and currently do not include baryons.

Finally, to extract cosmological information from the statistical properties 
of the density field means we require a covariance matrix. Nonlinear 
growth of dark matter structure will already correlate estimates of the matter power spectrum 
on different scales (e.g. Kiessling, et al. 2010), and including baryonic physics we can expect even stronger 
covariances between scales due to feedback processes.

The cosmic shear power spectrum 
depends on the matter power spectrum through an integral
over the line of sight distance, with a geometric lensing
kernel. The lensing power spectrum contains cosmological information,
through the lensing kernel, even on small scales. Rather than throw
this information away, we propose that one can include even very
uncertain non-linear scales in a lensing analsysis, if one correctly
marginalizes over the uncertainty. 
Using the path-integral marginalization techniques, presented in
Taylor \& Kitching (2010) and 
Kitching \& Taylor (2010), we derive an expression, in the
self-calibration regime, for the cosmic shear covariance that includes
the geometric lensing kernel information from all scales. 

If an informative upper bound on the functional behaviour of the
non-linear power spectrum can be determined 
(from simulations for example) the residual uncertainty in the functional
form of the power spectrum can be marginalized over simultaneously
with the cosmological parameters of interest. 
We place a requirement on the external fuctional prior 
such that the cosmological constraints
from future all-sky cosmic shear experiments are not degraded below a
level needed to determine dark energy physics.  

This article is organised as follows, in Section \ref{Method} we 
present the marginalized likelihoods and Fisher matrices for
the non-linear tomographic cosmic shear power spectrum. In Section
\ref{Results} we present results, and in Section
\ref{Conclusion} we present our conclusions. 

\section{Method}
\label{Method}

\subsection{Cosmic Shear Tomography}
In this article we will focus on tomographic cosmic shear, where the
shape information of galaxies is used, and objects are binned by their
estimated redshift, to create a series of 2D cosmic shear maps. 
The auto- and cross-correlation of these maps can be used to
generate a series of power spectra 
 \be
 \label{GG}
 C_{ij}(\ell)=\int_0^{r_H}\! {\rm d}r \, W^{\rm GG}_{ij}(r)
 P_{\delta\delta}\!\left(\frac{\ell}{S_k(r)};r \right),
 \ee
where lensing weight can be expressed as
 \be
 W^{\rm GG}_{ij}(r)=\frac{q_i(r)q_j(r)}{S^2_k(r)},
\ee
and the kernel is 
\be
q_i(r) = \frac{3 H^2_0 \Omega_m S_k(r)}{2 a(r)}
\int_r^{r_H}\!  dr' \, p_i(r')
\frac{S_k(r'-r)}{S_k(r')}.
\ee
We follow the notation of Joachimi \& Bridle (2009).  The
comoving distance is $r$, $r_H$ is the horizon distance, while
$S_k(r)=\sin(r), r,\sinh(r)$ for curvatures $k=-1,0,+1$, $a$ is the scale factor
and $P_{\delta\delta}(\ell/S_k(r); r)$ is the 3D density  
matter power spectrum. The comoving galaxy probability 
distribution is given by $p_i(r)$. The
$ij$ subscripts refer to redshift bins, where the shear field is
approximated as a series of correlated 2D planes. This can be
generalized to a full 3D shear estimator (Kitching, Heavens, Miller, 2010).
We will neglect intrinsic alignment terms so that the covariance
of the cosmic shear can be written as 
\ba
\label{ccc}
\C_{\mu\nu}(\ell)=2C_{jm}(\ell)C_{il}(\ell)
\ea
where $\mu={ij}$ and $\nu={ml}$ refer to redshift-bin pairs. 

\subsection{Information in the Non-Linear Regime}

\begin{figure}
  \includegraphics[angle=0,clip=,width=\columnwidth]{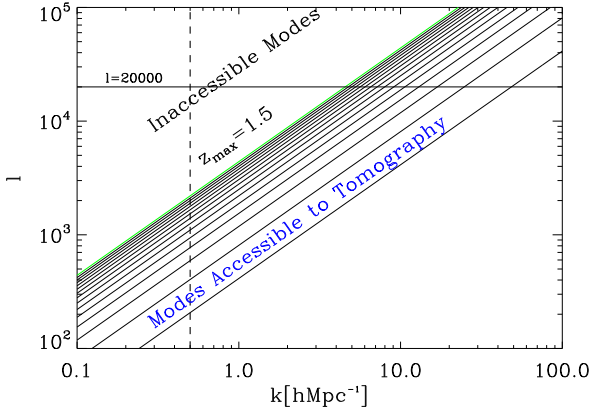}
 \caption{The $k$ modes that are \emph{in principle}
   accessible to a tomographic experiment. As an
   example we use a $15$ bin tomography experiment with bins at $0.1n$
   from $z=0.1$ to $1.5$; solid lines drawn are $\ell=kr(z)$.
   Tomography uses modes that lie on the lines
   shown. 3D cosmic shear (Kitching, Heavens, Miller, 2010) would use
   all modes below the green line, which marks the maximum redshift of
   the example experiment. The dashed line shows the approximate
   non-linear cut off of $k\approx 0.5h$Mpc$^{-1}$.}
 \label{vbias0}
\end{figure}

One subtlety, a result of the Limber
approximation used in equation (\ref{GG}), and explicitly highlighted by Kitching, Heavens \&
Miller (2010), is that the radial $k$-modes in the matter power
spectrum are associated with the
azimuthal $\ell$-modes through $\ell=k r(z)$ where $r(z)$ is the
comoving distance for redshift $z$. 
Commonly angular wavenumbers of up to $\ell_{\rm max} = 20$,$000$ 
are used in theoretical predictions of weak lensing power. 
If no cut in Fourier 3D wavenumber $k$ is applied for large $\ell$ modes the 
resulting scales probed in the matter power spectrum can extend into 
the very highly non-linear regime. For instance for $\ell_{\rm max} = 20$,$000$ and 
$r(z=0.1)=400$Mpc$/h$ we find $k=70h$Mpc$^{-1}$, or a physical scale of $126$Kpc$/h$, 
within the dark matter halo of galaxies. We demonstrate this in Figure \ref{vbias0} where we show
the ($\ell$, $k$) plane and the accessible modes, that lie on the lines
for a $15$-bin tomography experiment from redshift $0.1$ to $1.5$; it is
clear that all modes with $\ell\gs 2000$ probe the non-linear regime
exclusively.

In this article we will make defined cut in radial $k$-mode, where
$k\leq k_{\rm max}$
for all tomographic bins. This allows scales to be probed in a 
controlled manner so that the non-linear or very non-linear regimes can be
removed from the analyses. Equivalently, this can be intepreted as a
redshift-dependent $\ell$ cut where $\ell_{\rm max}(z)=k_{\rm max}r(z)$.    

\begin{figure}
  \includegraphics[angle=0,clip=,width=\columnwidth]{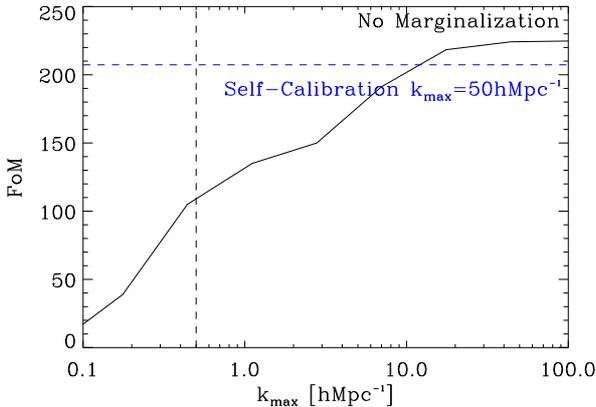}
 \caption{The dark energy Figure of Merit (FoM) as a function of the
   maximum $k$-mode used in the tomographic cosmic shear Fisher
   matrix analysis for a Euclid-like survey, with no external priors
   included. The maximum $\ell$ range in the $C(\ell)$ power
   spectrum is calculated as a function of redshift using
   $\ell_{\rm max}(z)={\rm min}[k_{\rm max}r(z),2\times 10^5]$.}
 \label{vbias1}
\end{figure}

In Figure \ref{vbias1} we show the predicted dark energy Figure of
Merit\footnote{
FoM; Albrecht et
al., (2006), defined as the area constrained in the uncorrelated 
($w_0$,$w_a$) plane with
${\rm FoM}=1/[\sqrt(F^{-1}_{w0w0}F^{-1}_{wawa}-(F^{-1}_{w0wa})^2)]$.}, calculated 
using a Fisher matrix (Tegmark, Taylor \& Heavens, 1997; see Hu 1999 for the 
cosmic shear tomography Fisher matrix), as a function
of the maximum $k$-mode used in the matter power spectrum for a
Euclid-like\footnote{Refregier et al., (2010), $20,000$ square degrees, with median
redshift of $z_m=1.0$, $35$ galaxies per square arcminute, 
a photometric redshift uncertainty of $0.03(1+z)$. We assume the
galaxy number density is given by  $n(z)\propto z^2{\rm
exp}(-1.4z/z_m)^{1.5}$, and use $10$ redshift bins. Throughout we
include no priors.} tomographic cosmic shear survey, and a set of 
cosmological parameters that includes curvature\footnote{Throughout
  we will use a cosmological parameter set that
allows for curved cosmologies with parameters $\Omega_m$,
$\Omega_{de}$, $\Omega_b$, $h$, $\sigma_8$, and $n_s$ given by
$(0.25,0.7,0.75,0.8,0.95)$, and parameterize the dark energy
equation-of-state using a first-order Taylor expansion,
$w(z)=w_0+(1-a)w_a$ (Linder, 2003; Chevallier and Polarski, 2001)
with $(w_0,w_a)=(-0.95, 0.0)$.}. 
It is clear that the majority of the dark energy information comes
from modes in the non-linear regime (although this statement contains some uncertainty because 
we have used the Smith et al, 2003 {\tt halofit} correction). 

We find convergence at $k_{\rm max}\approx
50h$Mpc$^{-1}$. This is in agreement with previous studies (e.g. Dore,
Tingting, Ue-Li, 2009). 
We can understand the behavior of this plot where 
the increase in FoM around $k=0.1h$Mpc$^{-1}$ 
corresponds to the nonlinear regime for the matter power-spectrum. 
The flattening between $k=0.5$ and $k=3h$Mpc$^{-1}$ occurs near the 
linear peak of the matter power spectrum which is less sensitive to dark energy. 
A second increase in information appears around the peak of the nonlinear regime, 
and then again saturates at very high $k$.

\subsection{Path Integral Marginalization}

The main issue we are concerned with in this paper is the uncertainty
in the shape of the nonlinear matter power spectrum due to baryon
effects. In the absence of an accurate estimate of the nonlinear
matter power spectrum one can either attempt to self-calibrate it from
the cosmic shear data itself, or apply an informative external prior
on its shape from hydrodynamical simulations. The self-calibration
approach finds the maximum likelihood solution to the free functional
nonlinear matter power spectrum using the shear data alone. For a
Gaussian likelihood for the matter power spectrum this can be done in
a single-step using Newton’s Method or a quadratic estimator (Taylor
\& Kitching 2010), or iteratively.  We then analytically marginalize
over all functional forms for the power spectrum permitted by the
data, again assuming Gaussian-distributed nonlinear power. The
resulting marginalized likelihood function for cosmological parameters
is now independent of any initial fiducial choice of the nonlinear
matter power, since we have found its maximum likelihood value. This
method can be applied to an arbitrary likelihood function for the
data, but if we have Gaussian-distributed data, the resulting
marginalized likelihood function for the data is again a Gaussian with
a new covariance matrix,
\be
 \label{lll11}
    \C^M = [\C^{-1} - \C^{-1}\Pp\C^{-1}]^{-1}
 \ee
where $\C$ is the orginal covariance and
 \be
    \Pp=\int \! dx' dx'' \,
    \frac{\delta \mub [\psi(x)]}{\delta  \psi_{\alpha}(x')}
    F_{\alpha\beta}^{-1}(x',x'')
    \frac{\delta \mub^\dag [\psi(x)]}{\delta  \psi_{\beta}(x'')},
    \label{flatprior}
 \ee
for functions $\psi_{\alpha}(x)$, 
and we assume the functional Fisher matrix, $F_{\alpha\beta}(x,x')$
for the function $\psi_{\alpha}(x)$, is invertible. $\mub$ is the mean 
signal, which in our case is the $C_{ij}(\ell)$ tomographic power 
spectrum.

If we assume an informative external prior, for example, if we know
something about how baryons will affect the matter power spectrum from
hydrodynamical simulations, we can include this information by
constraining its shape to some accuracy which in general will depend
on scale. This additional information can be included in our formalism
by multiplying the likelihood with a Gaussian prior with some
covariance $C_{\alpha\beta}(x',x'')$. After marginalization, we again find an analytic
expression for a general likelihood. In the case of
Gaussian-distributed data this also modifies the marginalized
covariance, $\C^M$. This in turn can be simplified using the Woodbury
relation to find that the marginalized data covariance is the original
covariance $\C$ with a new term added, 
\be
\label{csys}
\C_M = \C + \int dx' dx'' \,
C_{\alpha\beta}(x',x'') \frac{\delta
  \mub[\psi(x)]}{\delta
  \psi_\alpha(x')} \frac{\delta \mub^t[\psi(x)]}{\delta
  \psi_\beta(x'')}.
\label{margcov}
\ee 

In the following subsections we derive an
expression for the covariance of the tomographic cosmic shear power
spectrum in each case. Given the modified covariance $\C^M$ 
a likelihood function can then be constructed 
that accounts for unknown functional behaviour in the matter
power spectrum
 \be
 \label{mmml}
    \calL_0 = \Delta D_{\mu} [\C^M]_{\mu\nu}^{-1} \Delta D_{\nu}^t + \Tr \ln \C^M,
 \ee
where $\Delta D_{\mu}=C^{\rm theory}_{ij}(\ell)-C^{\rm
  data}_{ij}(\ell)$, which could be used in data analysis. We assume 
a Gaussian likelihood, but in fact any likelihood estimator can
be used (see Taylor \& Kitching, 2010). For predictive forecasts an 
associated Fisher matrix can be constructed assuming that the
cosmological information is the mean tomographic power spectrum at
each redshift 
\be
\label{margF}
F_{ab}=
\half \sum_{\mu \nu}\int \frac{\ell d \ell}{2 \pi}
\left(\left[\C^M_{\mu\nu}(\ell)\right]^{-1}
\frac{\partial C_{\mu}(\ell)}{\partial\theta_a}
\frac{\partial C_{\nu}(\ell)}{\partial\theta_b}\right)
\ee
for a set of cosmological parameters, $\btheta$.

\subsection{Self-Calibration}

In the self-calibration regime the cosmic shear data itself is used to 
measure the non-linear power spectrum simultaneously with cosmological 
parameters. To calculate the impact of this self-calibration we first find the 
functional derivative of the lensing tomographic power spectrum
with respect to the matter power sepctrum, given by 
 \be
\label{dddd}
 \frac{\delta C_{ij,j>i}(\ell)}{\delta P_{\delta\delta}(\ell'/r,r)}
  =W^{GG}_{ij,j>i}(r) \frac{ 2 \pi \delta_D(\ell-\ell')}{\ell}.
 \ee
For a flat-prior in function-space over the non-linear matter power
spectrum, where we use the data itself to fit and
marginalize over uncertainty in the 
matter power spectrum, the marginalized functional covariance
is given by 
\ba
\label{ffgghe}
\C^M_{\mu\nu}(\ell) = [\C^{-1}_{\mu\nu}(\ell) -
  \C_{\mu\mu'}^{-1}(\ell)\Pp_{\mu'\nu'}(\ell)\C^{-1}_{\nu'\nu}(\ell)]^{-1}
\ea
where  
\ba
\Pb_{\mu\nu}(\ell) = \int_0^{r_H} \!\! dr dr' \,
W^{GG}_{\mu}(r)[F(\ell; r, r')]^{-1} W^{GG}_{\nu}(r'),
\ea
and the functional Fisher matrix for the matter power spectrum is
\be
\label{ffgghf}
F(\ell;r,r')
=\sum_{\mu\nu}\C^{-1}_{\mu\nu}(\ell)W^{GG}_{\mu}(r)W^{GG}_{\nu}(r').
\ee
In practice this functional Fisher matrix is binned in redshift and scale
using $200$ bins in each dimension in this paper.

After marginalizing over the nonlinear matter power spectrum, there is
still cosmological information in the lensing power spectrum  through
the lensing kernel. The functional Fisher matrix in equation (\ref{ffgghf})
picks out the geometric dependency of the tomographic cosmic shear
power spectrum.  Marginalizing over the functional matter power
spectrum has some similarity to the geometric shear-ratio test (Jain \&
Taylor 2003, Taylor et al 2007, Kitching et al 2007), where the shear
signal behind galaxy clusters and groups is ratioed, and the mass
drops out leaving just the geometric part of the lensing signal. This
method can be used in the fully nonlinear clustering regime as there
is no sensitivity to clustering.  

\subsection{External Prior}

For an external prior we 
take the functional derivative 
of the tomographic cosmic shear
power spectrum with respect to the matter power spectrum (equation
\ref{dddd}) and, assumming a Gaussian distribution in function-space can now write a 
covariance for the tomographic cosmic shear power spectrum that
marginalizes over all unknown functional behaviour of the matter power
spectrum 
\ba
\label{margC2}
\C^M_{\mu\nu}(\ell)=\C_{\mu\nu}(\ell)+
\int_0^{r_H} \!\! {\rm d}r \,  \sigma^2_P(\ell/r,r) W^{GG}_{ij}(r)
W^{GG}_{lm} (r),
\ea
where $\sigma^2_P(\ell/r,r)=\lgl |\delta
P_{\delta\delta}(\ell/r,r)|^2\rgl$ is the functional scatter in
$P_{\delta\delta}$ and we have assumed the functional covariance is 
diagonal in $\ell$ and $r$.

\section{Results}
\label{Results}

Here we present cosmic shear Fisher matrix forecasts, marginalizing
over uncertainty in the matter power spectrum, for the
self-calibration and external prior cases described in Section
\ref{Method}.

\subsection{Self-Calibration}

In Figure \ref{vbias1} we show the FoM for the case that the matter
power spectrum is measured directly from the cosmic shear data, upto a
maximum $k=50h$Mpc$^{-1}$. We find that relative to the
non-marginalization case (where the power is assumed to be known
exactly) there is a reduction in the ability of the cosmic shear
survey to constrain dark energy parameters by $10\%$. This is 
equivalent to the maximum $k$-range being reduced 
from $k=50h$Mpc$^{-1}$ to $k\approx 10h$Mpc$^{-1}$.

We note that the information in the self-calibration regime comes from the geometric 
constraints on cosmological parameters that come from the 
cross-component tomographic bin combinations (for a single bin, 
with auto-correlation information only 
the marginalized covariance, equation \ref{ffgghe}, $\C^M\rightarrow\infty$). 
Because the self-calibration recovers the FoM to within 10\% of the unmarginalized case, 
this suggests that the majority of the dark energy information from 
cosmic shear tomography, in the non-linear regime, comes from the 
geometric part of the lensing kernel. 

\subsection{External Prior} 

In this section we will place requirements on the functional variance from an 
external prior on the non-linear power spectrum so that the expected dark energy
cosmological constraints remain unaffected. We parameterise the
functional variance using 
\be 
\label{44}
\frac{\sigma^2_P(k,r)}{P^2(k,z)}=a_0+a_1k \,\,\,\,\,\,\,\forall\,\,\,k>0.5h{\rm Mpc}^{-1}
\ee
where we will investigate a constant uncertainty and one that scales
linearly with the $k$-mode, here $k=\ell/r$ in equation
(\ref{margC2}). This is an extension of the type of path integral
marginalization used in Kitching \& Taylor (2010) where only constant
functional variances were considered, here we consider varying
functional variance. Note that we assume that the variance is not a
function of redhift, and that $a_1$ has units of $h^{-1}$Mpc.  
\begin{figure}
  \includegraphics[angle=0,clip=,width=\columnwidth]{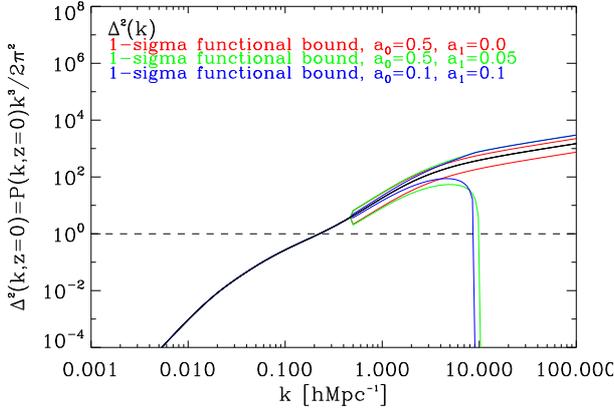}
 \caption{The dark matter power spectrum at a redshift of zero, as a
   function of scale. We show 3 examples of the type of $1$-sigma
   functional variances that can be considered using equation
   \ref{44}. Below $k=0.5h$Mpc$^{-1}$ we assume that the power
   spectrum is known exactly, which extends slightly into the
   non-linear regime. The horizontal
   dashed line marks $\Delta^2(k)=1$, which is where there is
   approximately a $1\%$ deviation from linearity. 
   For the calcuations in Section \ref{Results} that actual 
   values of $a_0$ and $a_1$ are an order of magnitude smaller than those shown here, 
   which are are illustrative purposes.}
 \label{vbias2}
\end{figure}
Figure \ref{vbias2} shows an example of the type of functional bounds that we
will consider (we show the dimensionless power
$\Delta^2(k)=P(k)k^3/2\pi^2$ so that the non-linear regime is 
marked by $\Delta^2(k)\gs 1$), throughout we assume that below $k=0.5h$Mpc$^{-1}$ the
power spectrum is known exactly (zero functional variance) and we take
an upper limit in wavenumber of $k\leq k_{\rm
  max}=50h$Mpc$^{-1}$. We use the Eisenstein \& Hu (1999)
linear power spectrum and the Smith et al. (2003) non-linear
correction as the fiducial function. 
\begin{figure}
  \includegraphics[angle=0,clip=,width=\columnwidth]{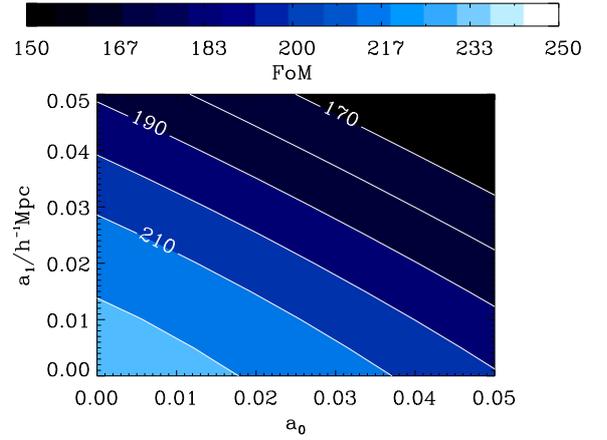}
  \includegraphics[angle=0,clip=,width=\columnwidth]{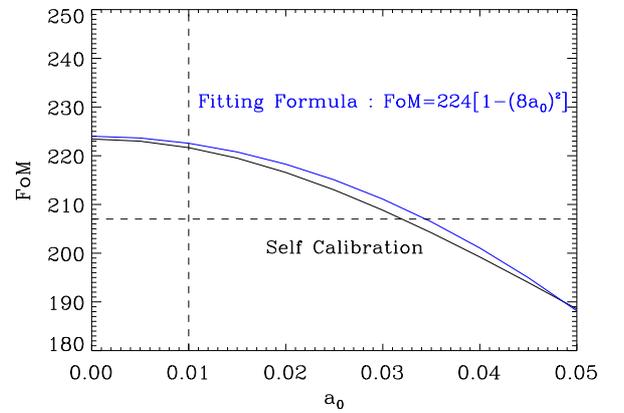}
 \caption{The FoM, for a Euclid-like tomographic cosmic shear survey alone, 
   as the functional variance of the external prior is varied, as parameterised by equation 
   (\ref{44}). The FoM degrades as the functional variance is
   increased. The upper panel shows how the FoM degrades as a 
   functional of $a_0$ and $a_1$, the lower panel show the dependency
   on $a_0$ keeping $a_1=0$. We provide a simple 
   fitting formula (blue, light gray, line) that relates a constant
   functional variance $a_0$ to the FoM. If the functional 
   variance is $\ls 0.01$ (vertical dashed line) 
   the FoM is degraded by less than $1\%$. The horizontal dashed line
   shows the FoM in the self calibration regime.}
 \label{vbias2a}
\end{figure}

In Figure \ref{vbias2a} we show how the FoM for a tomographic cosmic shear, for the survey outlined in Section \ref{Method}, 
changes as a function of the functional variance on the non-linear power spectrum from an external prior. We find that 
if the functional variance can be constrained to $\ls 1\%$ then the FoM degradation is at most $1\%$ relative to the case where 
no marginalization is performed. We provide a simple fitting formula for the dependency of the FoM on the 
constant contribution to the functional variance
\be 
\label{fit}
{\rm FoM}=224[1-(8a_0)^2],
\ee
which is valid for $a_0\ls 0.1$.

\section{Conclusion}
\label{Conclusion}

To conclude we find that the non-linear and baryon-dominated 
part of the matter power
spectrum contains, above wavenumbers of $k\gs 0.5h$Mpc$^{-1}$, 
half of the information content on dark energy parameters,
parameterised through the Figure of Merit. However the lack of
knowledge about this regime, and the complex simulated modelling needed to
correctly constrain its behaviour as a function of environment, scale
and cosmology means that the uncertainty on the non-linear power
spectrum must be correctly accounted for in cosmic shear
surveys. This article has some resonance with 
previous work, that
parameterise the uncertainty in the non-linear power spectrum and
marginalize over those parameters (Zhang et al., 2009; 
Rudd et al., 2008; Zenter et al., 2008; Huterer et al., 2006; Jing
et al., 2000) or attempt to modify the data to minimise the effect
(Huterer \& White, 2005), however all these assume parameterized 
models, that may not be able to reflect the real effect of baryons.

We have derived likelihood expressions for the tomographic cosmic
shear power spectrum in the cases that the functional matter power spectrum is
self-calibrated from the data itself, and in the case that an external
prior on the functional variation of the matter power spectrum is
available. 

\begin{figure}
  \includegraphics[angle=0,clip=,width=\columnwidth]{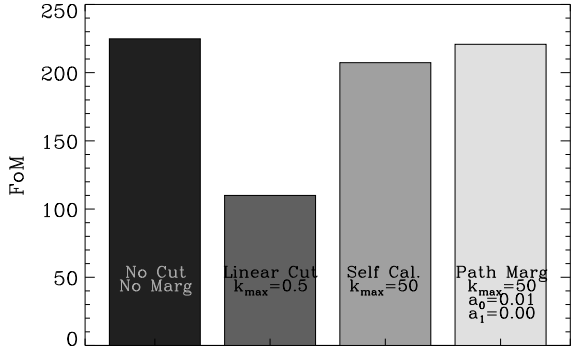}
 \caption{Summary of the main results. If the non-linear scales are
   removed using a hard cut in $k$-modes then the FoM can be reduced
   by $50\%$. In the functional self-calibration regime the reduction
   is less severe with a relative reduction of $10\%$, finally if a
   $1\%$ external prior on the functional variation of the non-linear
   power can be applied then the FoM is recovered. All $k$-modes are
   in units of $[h$Mpc$^{-1}]$. The parameters $a_0$ and $a_1$ refer
   the shape of the functional variance as a function of scale, given
   by equation (\ref{44}).}
 \label{vbias3}
\end{figure}

We summarise our results in Figure \ref{vbias3}. 
\begin{itemize}
\item 
With no external priors, a Euclid-like cosmic shear survey, with a 
$k_{\rm max}=50h$Mpc$^{-1}$ could
achieve a FoM $\approx 220$ from cosmic shear tomography alone.\\
\item 
If the non-linear matterpower spectrum is completely removed using a
hard cut in $k$-modes of $k_{\rm max}=0.5h$Mpc$^{-1}$ then the FoM is
reduced by a factor of $50\%$. \\
\item 
In the functional self-calibration regime the cosmic shear survey can recover the
FoM, with only a $10\%$ reduction in the FoM. \\
\item 
By including an informative prior, from simulations for example, the
orginal FoM can be recovered if the functional variation of the
non-linear matter power spectrum is known to $\sim 1\%$ to 
$k=50h$Mpc$^{-1}$, or a physical scale of $\sim 120$Kpc$/h$. 
\end{itemize}
Finally we note that in the self-calibration regime, the information
used to constrain the cosmological information, through the lensing
kernel, has some similarities with the shear-ratio method, 
where cluster scale weak lensing is isolated.

Constraining the non-linear power spectrum to 1\% functional accuracy down to $120$kpc$/h$ 
as a function of scale, redshift and cosmology is a significant theoretical and observational challenge. 
On the theoretical side modeling the baryons on the scale of galaxy clusters is already a challenge. 
Extending this to group and individual galaxy haloes will require a much deeper understanding of 
the baryonic processes on these scales. 
On the observational side, weak lensing itself, and galaxy-galaxy lensing, 
can provide much empirical information about the mass distribution which can be compared with 
stellar and gaseous components.
These are challenges that must be
realised if we are to 
fully exploit the potential of tomographic cosmic shear experiments.  
\\

\noindent{\em Acknowledgements:} TDK is supported by STFC
rolling grant number RA0888. We thank the eScience institute Edinburgh
for hosting a workshop on n-body simulations in cosmology. We thank
Alan Heavens, Catherine Heymans, Fergus Simpson 
for interesting discussions on this topic.



\onecolumn

\end{document}